\documentstyle[12pt]{article}

\author{Roustam Zalaletdinov\thanks{E-mail address:
zala@axp3g9.icra.it}\\[5mm]
{\em International Center for Relativistic Astrophysics, Departamento di
Fisica}\\
{\em Universit\'a di Roma "La Sapienza", P.le Aldo Moro 5, Roma 00185,
Italy}\\[2mm]
{\em Department of Theoretical Physics, Institute of Nuclear Physics}\\
{\em Uzbek Academy of Sciences, Tashkent 702132, Uzbekistan, C.I.S.}}
\title{\LARGE \bf Approximate Symmetries \\
in General Relativity}
\date{}

\begin{document}

\maketitle

\thispagestyle{empty}

\begin{abstract}
The problem of finding an appropriate geometrical/physical 
index for measuring a degree of inhomogeneity for a given
space-time manifold is posed. Interrelations with 
the problem of understanding the gravitational/informational entropy 
are pointed out. An approach based on the notion of approximate 
symmetry is proposed. A number of 
related results on definitions of approximate symmetries 
known from literature are briefly reviewed
with emphasis on their geometrical/physical content. A definition of 
a Killing-like symmetry is given and a classification
theorem for all possible averaged space-times acquiring Killing-like
symmetries upon
averaging out a space-time with a homothetic Killing symmetry is proved.
\end{abstract}

\newpage

\section{Entropy, Information, Inhomogeneity}
\label{sec:eii}
In modern gravitational physics there exist a number of proposed 
definitions of gravitational entropy. Amongst them one can mention
the black hole entropy by Bekenstein~\cite{Beke:1973} and
Hawking~\cite{Hawk:1975}
which relates the gravitational entropy with the black hole surface area,  
the space-time entropy by Penrose~\cite{Penr:1979} based on the idea of
employing
the Weyl tensor for `measuring' the pure gravitational content of a given 
space-time, Hu's cosmological entropy~\cite{Hu:1983} due to particle
production 
in anisotropically expanding Universe, Blanderberger-Mukhanov-Prokopec's
non-equilibrium entropy~\cite{BMP:1993} for a classical stochastic field
applied for 
density perturbations in an inflationary Universe, the intrinsic entropy
by 
Smolin~\cite{Smol:1985} measuring the irreversibility inherent in 
conversion any form of matter into gravitational radiation.
Despite relative consensus of opinions regarding the usefulness and
physical 
adequacy in applying the known definitions of gravitational entropy for 
the corresponding domains of gravitational phenomena, there is
still much controversy in understanding (formulating) the underlying 
foundations of classical and/or quantum gravitational physics that are 
generally expected to bring about a generic definition of the notion 
of gravitational entropy, 
it being expected to be of geometrical nature. Two aspects 
here are of the primary importance: (i) missing links and 
interrelations between different definitions; (ii) absence, in most cases,
of clear geometrical interpretations of the proposed notions of entropy.

The challenging problem of gravitational entropy can be considered in 
broader context as that of finding relations between the informational and 
gravitational entropy (see a discussion 
in~\cite{Beke:1973}, \cite{Beke:1994}, \cite{Davi:1978} and references
therein).
The manifold's entropy here is to be understood as a kind of geometrical 
entropy measuring the information encoded in a space-time manifold and 
being related with 
manifold's inhomogeneity. In such approach the key point (assumption) is
that an evolving 
gravitational system tends to a final symmetric (relatively) homogeneous 
state\footnote{An important class of self-organizing, 
dissipative physical systems~\cite{Hake:1978} is not considered here.} 
with subsequently increasing the system's entropy, or, its physical
homogeneity
which is reflected, in general, in the homogeneity of the space-time.
Indeed, a `structureless' smooth, highly symmetric homogeneous manifold
(i.e. 
when the 
system is disordered and requires less information to describe it) may be
considered to possess maximum entropy, while a lumpy inhomogeneous, highly
structured 
manifold (i.e. when the system is ordered and requires a great deal of
information to
describe it) does have a smaller value of entropy. 
In such context the entropy of a system measures one's uncertainty or lack 
of information about the actual internal configuration of the system. 
Given a set of system's states determined by probabilities $p_n$, the
system
entropy is defined due to Shannon's formula~\cite{Shan:1948} as
\begin{equation}
\label{shan}
S=-\sum_{n}p_n \ln{p_n}.
\end{equation}
New information $\Delta I$ about the system imposes some constraints on
the probabilities 
$p_n$, which results~\cite{Bril:1956} in a decrease $\Delta S$ in one's
uncertainty about 
the internal state of the system. Due to Brillouin's 
identification of information with negative entropy~\cite{Bril:1956} the 
property is formalized by the relation
\begin{equation}
\label{bril}
\Delta I=-\Delta S.
\end{equation}
With statistical mechanics following from the information 
theory~\cite{Jayn:1957a}, \cite{Jayn:1957b} and 
the second law of thermodynamics valid, the relations (\ref{shan}) and
(\ref{bril})
between information and negative
entropy  have proved to be useful in understanding 
some geometrical features of 
gravitational entropy, in particular, Bekenstein-Hawking's
entropy~\cite{Davi:1978}.

\section{Inhomogeneity and Approximate Symmetry}
\label{sec:ias}
Modern gravitational physics based on the pseudo-Riemannian geometrical
picture of a space-time manifold poses another problem of finding an 
appropriate geometrical/physical index for measuring a degree of
inhomogeneity 
for a given space-time manifold. The theoretical description of
gravitational field
based, as for all other known interactions, on the notion of symmetry,
i.e.
the metric tensor is considered to be known if 
one knows the corresponding group of the space-time manifold 
symmetries\footnote{All gauge freedom is 
always inside the symmetry group.}.
Intuitively, to conclude which manifold is {\em more} inhomogeneous one
has 
to compare their symmetry groups - the more symmetries 
a manifold has, the more homogeneous it is. It is implicitly 
assumed here that the measure of inhomogeneity is in the difference of the
two
symmetry groups. However, such an approach does not have any definite
means to 
`measure' inhomogeneity. Indeed, how to decide which manifold is more 
inhomogeneous - one having, say, a couple of Killing vectors or another
possessing only one covariantly constant vector? In practice, to 
construct an inhomogeneous manifold 
one usually takes a perturbed manifold with a metric 
$g_{\mu \nu}=g^{(0)}_{\mu \nu} + \epsilon h_{\mu \nu}$ with 
a presumed background manifold
$g^{(0)}_{\mu \nu}$ and the perturbation functions $h_{\mu \nu}$ where
$\epsilon \ll 0$
is the smallness parameter.
The notion of background here is central for it is
taken by definition as a smooth and homogeneous {\em reference} manifold.
It enables one to determine the above mentioned difference between 
two manifolds (the perturbed and background ones) through the 
inhomogeneous perturbations 
$h_{\mu \nu}$, the notion  
of background metric itself assuming a kind of 
smoothing procedure~\cite{Matz:1968a}, \cite{Mont-Zala:1999} usually taken
as a space, or space-time volume averaging. 

An index of inhomogeneity is expected to have non-trivial physical and
geometrical 
content, as well as being closely related with information and entropy.
Indeed, let us
assume that an index $\chi$ measures inhomogeneity of a 
gravitating physical system, 
i.e. that of its space-time, and let initially at $t_0$ 
the system have a symmetry group $G_0$, 
an inhomogeneity index $\chi_0$ and a characteristic inhomogeneity length
$l_0$. 
During evolution of the system to an equilibrium state  
for $\Delta t=t_1 - t_0$, its entropy increases for $\Delta S$ with a
decrease in 
information $\Delta I$ about its internal configuration 
(due to washing out system's initial conditions) in accordance with 
Brillouin's relation (\ref{bril}). At the same time the system becomes
more 
homogeneous, i.e. decreases for $\Delta \chi=\chi_1 - \chi_0$. Therefore,
at $t_1$
the system has a symmetry group $G_1$, $G_1 \subset G_0$, 
an inhomogeneity index $\chi_1$, $\chi_1 < \chi_0$,  
and a characteristic inhomogeneity length $l_1$, $l_1 \gg l_0$ and the 
following relation between information, entropy and inhomogeneity holds
\begin{equation}
\label{inho}
\Delta I=-\Delta S\sim \Delta \chi.
\end{equation}
The evolved state $\{G_1, \chi_1, l_1\}$ is coarse-grained compared with
the
initial one $\{G_0, \chi_0, l_0\}$ and the transition from the latter to
the former
may be accomplished by means of an appropriate smoothing operator with a 
smoothing scale $d_{\rm aver}$ such as $l_1 \gg d_{\rm aver} \gg l_0$. 
Through a sequence of more and more coarse-grained states $\{G_i, \chi_i,
l_i\}$ 
satisfying the relation (\ref{inho})
the system reaches eventually the equilibrium state $\{G_{\rm max}, 0 ,
L\}$,
i.e. the space-time manifold becomes homogeneous with the symmetry group
$G_{\rm max}$
and the vanishing inhomogeneity index, $\chi=0$, the characteristic
manifold length $L$ 
typically being manifold's curvature radius\footnote{The homogeneous
manifold may have 
several characteristic lengths like for a torus.}. 

The goal of this paper is to propose an approach to define an
inhomogeneity 
index by means of finding out an appropriate description of symmetries of 
inhomogeneous manifolds - the group $G$ from the above discussion. The
question 
is much finer that it seems at first sight. Indeed, such a `standard'
symmetry
as Killing's is likely to be `too symmetric and smooth' to serve for 
description of inhomogeities usually thought of as a kind of lumpiness or
ripples, i.e. an inhomogeneous manifold can be viewed as possessing
symmetry
that is not precise. The perturbation approach described above illustrates
this 
idea where the  
group of an inhomogeneous manifold is a distorted, or, approximate in some
sense, 
group of the background manifold, the former being taken as the direct
product 
of the latter by 
the 4-parametric gauge group of infinitesimal coordinate transformations.
The idea of approximate symmetry as relevant to the geometry of
inhomogeneous spaces
had been put forward first by Matzner~\cite{Matz:1968a}, \cite{Matz:1968b}
who had proposed a way of defining {\em almost Killing vectors}, a
generalization 
of the Killing symmetry (see Section \ref{sec:as}) useful for description
of 
spaces with gravitational radiation and related to an invariant definition
of
background. In a physical setting in the early sixties
Komar~\cite{Koma:1962}, 
\cite{Koma:1963} had introduced the concept of {\em semi-Killing vectors},
a kind of relaxed 
Killing symmetry (see Section \ref{sec:skv}), proved to be useful for the
formulation
of asymptotic covariant conservation laws for gravitational radiation. 
It should be pointed out that the problem of defining approximate
symmetries 
in a rigorous way is not simply academic because of its primary importance 
in modern cosmology for 
understanding the structure and evolution of our Universe with a hierarchy
of 
physical scales from stars to clusters of galaxies~\cite{Elli:1984}. 
The Universe is certainly 
inhomogeneous on smaller scales becoming smooth (averaged) for its largest
scale  
where its space-time geometry possesses isometries in accordance with the
Friedmann-Lema\^\i tre-Robertson-Walker cosmological model. 

In this paper the definitions of approximate symmetries 
known from literature are briefly reviewed
with emphasis on their geometrical/physical content and with pointing out 
corresponding
candidates for the inhomogeneity indexes. A general
definition of a Killing-like symmetry~\cite{Tava-Zala:1994},
\cite{Zala:1999} 
is given and a classification
theorem for all possible averaged space-times acquiring Killing-like
symmetries upon
averaging out a space-time with a homothetic Killing symmetry is proved.

\section{Approximate Symmetry Groups}
\label{sec:asg}
The main idea of the approach by Spero and 
Baierlein~\cite{Sper-Baie:1977}, \cite{Sper-Baie:1978} is to construct a 
3-parameter simply-transitive {\em approximate symmetry group} of an
inhomogeneous 
space-time metric\footnote{A space-time is said to be spatially
homogeneous if it is 
invariant under a 3-parameter abstract Lie group $G_3$ acting simply 
transitively on a family of spacelike hypersurfaces.} 
by minimizing a specific 3-volume-average of deviations 
of the orthonormal tetrad rotation coefficients from those in a Bianchi
type. 

Given a subset $\cal U \subset \cal S$, where $\cal S$ is a 
spacelike hypersurface with a metric $g_{ab}$ of a space-time with a
metric $^{4}g_{\mu \nu}$, one wishes to find a triad of orthonormal 
vectors\footnote{Capital Latin letters stand for the triad components and
the indexes run from 1 to 3.}
$\{ \mbox{\bf e}_A  \}$, $g_{ab} e^a_A e^b_B = \delta^A_B$,  
in $\cal U $ such that the commutation
coefficients
\begin{equation}
\label{rota}
\gamma^C_{AB} = g_{ab} [\mbox{\bf e}_A, \mbox{\bf e}_B]^a \mbox{\bf
e}^{bC}
= 2 e^a_{[A} \nabla_a e^b_{B]} e^C_b
\end{equation}
are as close as possible to some set of structure constants $C^C_{AB}$. 
The simply-transitive group 3-parameter Lie group to which $C^C_{AB}$ 
corresponds is said to be the approximate symmetry group of $g_{ab}$ and
$\{ \mbox{\bf e}_A  \}$ is said to be the best fit triad. 
This is done by requiring that $\{ \mbox{\bf e}_A  \}$ and $C^C_{AB}$ 
of the following 3-volume average of rotation coefficient deviations:
\begin{equation}
\label{I}
I \equiv \frac{1}{V} \int_{\cal U} \Delta^C_{AB} \Delta^C_{AB} dV 
+ 8 \lambda_A n_{AB} a_B + 2 \lambda_{[AB]} n^{[AB]}
\end{equation}
give a global minimum $\delta I=0$ under variation with respect to the
$e^a_A$.
Here the deviations are $\Delta^C_{AB} = \gamma^C_{AB} - C^C_{AB}$, 
the Lagrange multipliers $\lambda_A$ and $\lambda_{[AB]}$ ensure that
$C^C_{AB}$
are the structure constant antisymmetric in $A$ and $B$ and satisfying the 
Jacobi identities, the tensor density $n^{AB} = \frac{1}{2} C^{(A}_{CD} 
\epsilon^{B)CD}$ and the vector $a_B = \frac{1}{2} C^D_{BD}$ appear
in the irreducible decomposition of the structure constants
$C^C_{AB} = \epsilon_{ABD} n^{DC} + 2 a_{[A} \delta^C_{B]}$ and $V =
\int_{\cal U} dV$
is the 3-volume of the subset $\cal U$ with the 3-volume invariant measure
$dV$ .

The approach has enabled to reveal that the simply transitive group $G_3$
in Bianchi types 
VI$_h$, VII$_h$, VIII and IX are {\em stable} symmetries, i.e. a
homogeneous 
metric with one of these symmetry group types will preserve this group
type
when perturbed by arbitrary small metric perturbations. That means that
all the perturbed types possess the approximate symmetry group. 
The $G_3$ in all other types are unstable. All two-dimensional metrics and 
some three-dimensional metrics have been also analyzed. It is shown, in
particular,
that the approximate symmetry group of the 3-dimensional Kantowski-Sachs
space-time 
is the Bianchi type I and those of the Gowdy $T^3$ space-time are types I
or VI$_0$.

In the framework of the approach $I$ and $\lambda^2$ measure the degree of 
metric's inhomogeneity.
$I = 0$ and $\lambda^2 = 0$ iff the metric is homogeneous. $I > 0$ iff 
the metric is inhomogeneous, the greater $I$ the more inhomogeneous being
the 
metric, and vice-versa. The $I$ may be thought of as measuring the
`perpendicular
distance' in the 3-metric superspace from the metric under study to the 
nearest submanifold of homogeneous metrics. The role of $\lambda_A$ is
more 
subtle. No inhomogeneous solutions to the Einstein equations 
have been found~\cite{Sper-Baie:1978} in which $\lambda_A = 0$.

Thus, the approach allows one to find a {\em group} of approximate
symmetry
and the corresponding best fit tetrad may therefore be used to write down 
Einstein's equations explicitly. The technique is global apart possible 
difficulties with topology, and the approach has a 
sufficiently clear geometrical picture as far as Bianchi geometries are
involved.
At the same time, generalization to other geometries meets difficulties,
for the
technique depends drastically on the slicing algorithm of the space-time
under study,
and there are no links to other approaches to approximate symmetries. The
physical
meaning of the candidates for inhomogeneity indexes $I$ and $\lambda^2$ is 
not clear.

\section{Semi-Killing Vectors}
\label{sec:skv}
The concept of the semi-Killing symmetry 
appeared~\cite{Koma:1962}, \cite{Koma:1963} 
in searching conserved quantities for physically important space-times
(such as those with gravitational radiation) that do not have isometries.
The conservation of energy, momentum and angular momentum in
Lorentz-covariant field
theories in Minkowski space-time is well known to follow by requiring 
the invariance of physical laws with respect to infinitesimal 
transformations $\xi^\alpha(x^\mu)$ of the coordinate 
surfaces
\begin{equation}
\label{rigi}
x^{\prime \alpha} = x^\alpha + \xi^\alpha
\end{equation}
where $\xi^\alpha(x^\mu) = a^\alpha + \Omega^\alpha_\beta b^\beta$, with 
$a^\alpha$, $\Omega^\alpha_\beta$ and $b^\beta$ being rigid shift vector,
rotation
matrix and rotation parameters. Equivalently, one can start requiring 
the
invariance with respect to (\ref{rigi}) simultaneously with  
{\em the form-invariance} of the Minkowski metric tensor $\eta_{\mu \nu}$
with respect 
to the transformations, i.e. vanishing Lie-derivative 
of the metric $\pounds_{\xi} \eta_{\mu \nu} = 0$ that leads to Killing's
equation
for the vector $\xi^\alpha$:
\begin{equation}
\label{kill}
\xi_{\alpha ; \beta} + \xi_{\beta ; \alpha} = 0.
\end{equation}
The Killing vectors 
$\xi^\alpha$ are then coordinate-hypersurface orthogonal, to fix 
rigidly the preferred Minkowski coordinates in a flat space-time 
and single out the 
preferred conserved quantities, otherwise there are infinitely many
conserved quantities. In general relativity the infinitesimal
shift vector $\xi^\alpha$ is always a space-time function, so the 
question is how to {\em single out} a preferred system of curvilinear
coordinates
to get appropriate conserved quantities. In case of existing non-trivial
Killing vector fields (\ref{kill}) they had been 
shown~\cite{Koma:1962}, \cite{Trau:1962} to be appropriate 
in order to obtain the analogs of the preferred conserved quantities 
of Lorentz-covariant theories mentioned above. 

In general, however, Killing vectors cannot be found in many physically 
important space-time. From an analysis~\cite{Koma:1962}, \cite{Koma:1963} 
of the asymptotic properties of radiative solutions it has been suggested
to use a semi-Killing vector field $\xi^\alpha$ defined
by 
\begin{equation}
\label{semi}
\xi^\alpha (\xi_{\alpha ; \beta} + \xi_{\beta ; \alpha}) = 0, \quad
\xi^\alpha_{; \alpha} = 0,
\end{equation}
which is precisely half of the Killing vector, for finding a preferred
time
translation. If the vector field is timelike and spacelike-hypersurface 
orthogonal, the corresponding generalized energy-momentum flux vector of 
gravitational field $E^\alpha (\xi)) = -2 \xi^\beta R^\alpha_\beta$ is
conserved,
$E^\alpha (\xi)_{; \alpha} = 0$. The total generalized energy defined as 
\begin{equation}
\label{ener}
E(\xi) = \frac{1}{2 \kappa} \int E^\alpha dS_\alpha 
= \frac{1}{2 \kappa} \int E^4 (-\det g_{\mu \nu})^{\frac{1}{2}} d^3x
\end{equation}
can be shown~\cite{Koma:1962} to be positive-definite (here $\kappa$ is 
Einstein's constant). It is important to 
note that introduction of the semi-Killing vectors has been motivated
by the observation of Peres~\cite{Pere:comm} that if a coordinate system 
is chosen so that the hypersurfaces of constant time are minimal, 
and if the energy is determined by the strength of $1/r$ in the asymptotic 
behaviour of the $g_{00}$ metric component, then the energy is necessarily
positive-definite. Indeed, the coordinate hypersurfaces orthogonal to
a timelike semi-Killing vector are minimal, i.e. a 
solution to the 
Plateau problem - this is due to the first Eq.~(\ref{semi}), as well as 
harmonic due to the second Eq.~(\ref{semi})\footnote{Note that the
coordinate 
hypersurfaces of the Minkowski coordinates of flat space-time are 
both minimal and harmonic.}.

In the framework of the approach the total energy (\ref{ener}) serves 
as a measure of inhomogeneity of a space-time with a semi-Killing 
symmetry compared with that with isometries. The vanishing total 
energy $E(\xi) = 0$ implies that the space-time is locally flat. 

A definite advantage of the notion of semi-Killing vectors is their
transparent physical meaning and motivation and sufficiently clear
geometrical meaning, though it is not well-understood which limitations
are posed 
on a space-time by the conditions (\ref{semi}), especially by the second 
equation. At the same time, the approach is essentially local,
it does not have any generalizations and a symmetry group picture is not
available. Links to other approximate symmetry approaches are not known
except the almost symmetry (see Section \ref{sec:as}).

\section{Almost Symmetry}
\label{sec:as}
The almost symmetry of Matzner~\cite{Matz:1968a}, \cite{Matz:1968b} 
defines a measure $\lambda(\xi)$ of deviation from the Killing 
symmetry as a minimum of the specific functional
\begin{equation}
\label{almo}
0 \leq \lambda [\xi] = \frac{\int \xi^{(\alpha ; \beta)} \xi_{(\alpha ;
\beta)} dV}
{\int \xi^\alpha \xi_\alpha dV}, \quad dV \equiv (\det g_{\mu
\nu})^{\frac{1}{2}} d^4x
\end{equation}
where $\xi^\alpha$ is an arbitrary vector field. For the positive-definite
metrics, 
$\lambda$ is zero iff $\xi^\alpha$ is Killing. The differential equation
resulting from the variational problem $\delta \lambda = 0$ for
(\ref{almo}) reads
as an eigenvalue problem\footnote{The equation (\ref{eige}) can be viewed
as a
generalization of the equation $\xi^{(\alpha ; \beta)}{}_{; \beta} = 0$ by
Yano 
and Bochner~\cite{Yano-Boch:1953}, which in its turn seems to be another
generalization
(approximate symmetry) of Killing's equation (\ref{kill}). 
It should be pointed out that it is only true 
for the indefinite metric spaces, such as Lorentzian ones, for the second
order 
Yano-Bochner equation had been shown to be {\em equivalent} to Killing's
in 
positive-definite metric spaces.}
\begin{equation}
\label{eige}
\xi^{(\alpha ; \beta)}{}_{; \beta} + {_m}\lambda \xi^\alpha = 0
\end{equation}
with smallest eigenvalue $_0\lambda$ for positive-definite spaces from all
$_m\lambda$.
The upper bound for the $_0\lambda$ can be shown~\cite{Matz:1968b} 
to be the averaged curvature
$_0\lambda \leq \langle R \rangle = L^{-2}$ where $L$ is a typical length
of the 
problem. It is important to note that estimates of this type hold for the
eigenvalue 
$_0\lambda$ in {\em any} space and the idea of almost symmetry itself
enters when it 
turns out that $_0\lambda \ll L^{-2}$ . It means geometrically that an
almost
symmetry can be viewed as a kind of a deformation of a space with
isometries - what
one may call `inhomogenization' of an initially smooth space. 

The physical meaning of the almost symmetry becomes more clear while
considering
its application to high-frequency gravitational 
waves~\cite{Isaa:1968a}, \cite{Isaa:1968b}. Let $g_{\mu \nu}$ be a vacuum
metric 
which admits a steady coordinate system~\cite{ZTE:1996} such that the
metric can be written 
\begin{equation}
\label{stea}
g_{\mu \nu}=g^{(0)}_{\mu \nu} + \epsilon h_{\mu \nu} 
\end{equation}
with a background metric 
$g^{(0)}_{\mu \nu}$ and the perturbation functions $h_{\mu \nu}$ and 
the smallness parameter $\epsilon \ll 0$. Here $g^{(0)}_{\mu \nu}$ is a
slowly
varying function of space-time coordinates, $g^{(0)} = {\cal O} (1)$, 
$\partial g^{(0)} = g^{(0)} / L$, 
$h_{\mu \nu}$ is a rapidly varying function, $h = {\cal O} (l / L)$, 
$\partial h = h / l$, where 
$l$ is the short wavelength of radiation, $L$ is a typical background
length
(curvature radius) and due to the high-frequency approximation adopted 
$\epsilon \leq l / L$~\cite{Isaa:1968a}, \cite{MTW:1973}. Expansion of 
the Ricci tensor $R_{\mu \nu}$ for $g_{\mu \nu}$
in a series in powers of $\epsilon$ can be 
shown~\cite{Isaa:1968a}, \cite{MTW:1973} to bring about  
a certain propagation equation for $h_{\mu \nu}$, $R^{(1)}_{\mu \nu} = 0$, 
in the first order
and the equation $R^{(0)}_{\mu \nu} = - {\epsilon}^2 R^{(2)}_{\mu \nu}$ 
in the second order. To extract the part of $R^{(2)}_{\mu \nu}$ smooth
on the scale $l$, its space-time averaging over several 
wavelengths must be carried 
out~\cite{Bril-Hart:1964}, \cite{Isaa:1968b}, \cite{MTW:1973},
\cite{Zala:1996},
which gives Isaacson's equation
\begin{equation}
\label{isaa}
R_{\alpha \beta }^{(0)}(g^{(0)})=-\kappa T_{\alpha \beta}^{({\rm GW})}.
\end{equation}
where $T_{\alpha \beta }^{({\rm GW})}$ is Isaacson's energy-momentum 
tensor for gravitational waves
\begin{equation}
\label{t-gw}
T_{\alpha \beta }^{({\rm GW})} 
\equiv \frac{\epsilon^2}\kappa \langle R_{\alpha \beta }^{(2)} \rangle 
= \frac{\epsilon^2}{4\kappa } \langle h^{\mu \nu }{}_{;\alpha }h_{\mu \nu
;\beta } \rangle.
\end{equation}

Upon assuming the almost symmetry vector $\xi^\alpha$ to be only slowly
varying,
$\xi = {\cal O} (1)$, $\partial \xi = \xi / L$, calculation of the
integrand of 
the functional (\ref{almo}) gives
\begin{equation}
\label{inte}
\begin{array}{rcl}
4 \xi^{(\alpha ; \beta)} \xi_{(\alpha ; \beta)} & = & g^{(0) \alpha \mu}
g^{(0) \beta \nu}
\bigl[ \pounds_\xi g^{(0)}_{\alpha \beta} \pounds_\xi g^{(0)}_{\mu \nu}
\\[8pt]
& & + 2 \epsilon h_{\alpha \beta , \sigma} \xi^\sigma \pounds_\xi
g^{(0)}_{\mu \nu} 
+ \epsilon^2 h_{\alpha \beta , \sigma} \xi^\sigma h_{\mu \nu , \rho}
\xi^\rho 
+ {\cal O} (l) \bigr].
\end{array}
\end{equation}
With taking into account the denominator 
$\int \xi^\alpha \xi^\beta g^{(0)}_{\alpha \beta} (- \det g^{(0)}_{\mu
\nu})^{1/2} d^4 x$,
the first term in the right-hand side of (\ref{inte}) yields a number 
$4 \lambda_{g^{(0)}}[\xi]$ depending only on $\xi^\alpha$ and
$g^{(0)}_{\alpha \beta}$. 
In estimating 
the functional (\ref{almo}) for other terms in the right-hand side of
(\ref{inte}) 
one can observe that the fast-varying second term does not contribute into
the 
integral, and the only contribution coming from the third terms is due to 
its slowly varying part, 
$\int \epsilon^2 \langle g^{(0) \alpha \mu} h_{\alpha \beta , \sigma} 
g^{(0) \beta \nu} h_{\mu \nu , \rho} \rangle \xi^\sigma  \xi^\rho 
(- \det g^{(0)}_{\mu \nu})^{1/2} d^4 x$, the 
integrand of which contains the Isaacson's energy-momentum 
tensor for gravitational waves (\ref{t-gw}). As a result, $\lambda [\xi]$
for 
high-frequency radiation has been shown~\cite{Matz:1968b} to be of the
form
\begin{equation}
\label{hifr}
\lambda [\xi] = \lambda_{g^{(0)}}[\xi] + \lambda_{\rm rad} [\xi]
\end{equation}
where 
\begin{equation}
\label{radi}
\lambda_{\rm rad} [\xi] 
= \frac{ \kappa \int T_{\alpha \beta }^{({\rm GW})} \xi^\alpha \xi^\beta 
(- \det g^{(0)}_{\mu \nu})^{1/2} d^4 x}{\int g^{(0)}_{\alpha \beta }
\xi^\alpha \xi^\beta
(- \det g^{(0)}_{\mu \nu})^{1/2} d^4 x},
\end{equation}
both terms in (\ref{hifr}) being independent of $l$. 

In the almost symmetry approach $\lambda [\xi]$ (\ref{almo}) stands for a 
measure of inhomogeneity of a space with almost symmetry 
compared with that with isometries. 
As expected from the integral definition (\ref{almo}) for $\lambda [\xi]$
and from the estimation of its upper bound (see above), it is sampling 
the large-scale curvature of an almost symmetric space. Indeed, 
the estimation of $\lambda [\xi]$ of a space-time filled with
high-frequency radiation 
(\ref{hifr}) shows this explicitely since $\lambda_{g^{(0)}}[\xi]$ is
bounded
by the background curvature and $\lambda_{\rm rad} [\xi]$ being a kind 
of effective energy of radiation (\ref{radi}) is also a curvature in 
the background metric through the energy-momentum tensor 
$T_{\alpha \beta }^{({\rm GW})}$ and (\ref{isaa}). The $\lambda [\xi]$
given by 
(\ref{hifr}) does therefore measure only the large-scale space-time
curvature 
with a contribution from averaged ripples, yielding a total estimate for 
inhomogeneity of the space-time with high-frequency radiation compared
with 
a space-time $g_{\mu \nu}=g^{(0)}_{\mu \nu}$ with a Killing vector
$\zeta^\alpha$, 
$\pounds_\zeta g^{(0)}_{\mu \nu} = 0$. It is important to note that if an
almost symmetry vector $\xi^\alpha$ is timelike then $\lambda_{\rm rad}
[\xi]$ has a 
definite sign, it is positive (negative) if a timelike vector is positive
(negative) 
definite, for the energy density of high-frequency radiation is always 
positive-definite~\cite{Isaa:1968b}, 
$T_{\alpha \beta }^{({\rm GW})} \xi^\alpha \xi^\beta \geq 0$. For the 
$\lambda [\xi]$ itself (and $\lambda_{g^{(0)}}[\xi]$ in the above example)
its sign-definiteness can be only shown~\cite{Matz:1968b} for the
positive-definite 
metric spaces by finding a lower bound for $\lambda [\xi]$ so that 
$\lambda [\xi] \geq 0$.

One of the advantages of almost symmetry approach is its clear geometrical
meaning 
and the quantity $\lambda$ possesses also a specific physical
interpretation for 
physically interesting spaces. It should be pointed out that almost
symmetry has 
relevance~\cite{Matz:1968b} to semi-Killing vectors (see Section
\ref{sec:skv})
and almost-Killing vectors (see Section \ref{sec:aks}).
Though the technique is global and 
well-developed for positive-definite metrics, its meets difficulties in
generalization
to Lorentzian signature spaces. Also no symmetry group formulation of
almost
symmetry is known.

\section{Almost-Killing Vectors}
\label{sec:aks}
The notion of almost-Killing vectors has been introduced by
York~\cite{York:1974} 
in searching for `natural' vector fields in an asymptotically flat
space-time. A
general equation of this kind reads~\cite{Taub:1978}
\begin{equation}
\label{alki}
\xi_{\beta}{}_{; \alpha}{}^{; \alpha} + \xi^{\alpha}{}_{; \beta \alpha} 
- c \xi_{\alpha}{}^{; \beta}{}_{; \alpha} = 0.
\end{equation}
When $c = 0$ the equation (\ref{alki}) is called the almost-Killing 
equation\footnote{This is in fact Yano-Bochner's equation, see Section
\ref{sec:as}.}, 
and when $c = 1/2$ it is called the conformal almost-Killing one. There
are basically two 
observations regarding (\ref{alki}) which make interesting studying the
equation
in the Kerr space-time~\cite{Taub:1978}. 
Firstly, for $c = 2$ the equation reduces to Maxwell's equations for 
a source-free, test electromagnetic field with the vector 
potential $\xi^\alpha$ in the Kerr background. The equation is well-known
to
admit a remarkable decoupling of components and variables when being
solved
employing the Newman-Penrose formalism~\cite{Chan:1983}. Secondly,
compared with
the Killing equation (\ref{kill}) which has 16 components for 4 unknowns
and
therefore no solution in a general space-time, the equation (\ref{alki})
has 4 components for 4 unknowns and hence it is in general solvable.  

York's almost-Killing equation is obtained by acting on the Killing
equation with 
an additional covariant derivative. Any solution of Killing's equation is
also 
a solution of the almost-Killing equation,
and, generally, any asymptotic Killing 
vector\footnote{An asymptotic Killing vector in the Kerr space-time is
defined 
as a solution of the equation 
$\xi_{\alpha ; \beta} + \xi_{\beta ; \alpha} = {\cal O}(1/r^2)$ where 
$r$ is the Boyer-Lindquist radial coordinate~\cite{Boye-Lind:1967}.} 
is asymptotic to a solution of the 
almost-Killing equation. Thus the almost-Killing symmetry gives a natural
way 
of extending symmetries, whether approximate or exact, from infinity to
the 
entire space-time. The integral lines of a solution of the almost-Killing
equation, 
four mutually commuting, linear independent almost-Killing vectors, may be
used as a
`natural' Kerr asymptotic coordinates and serve as a coordinate grid
throughout
the entire Kerr space-time.

The situation is analogous with the conformal almost-Killing equation
which is 
a once covariantly differentiated version of the conformal Killing
equation
\begin{equation}
\label{ckv}
\xi_{\alpha ; \beta} + \xi_{\beta ; \alpha} 
= \frac{1}{2} g_{\alpha \beta} \nabla_\gamma \xi^\gamma.
\end{equation} 
Again, any Killing vector is also a solution of (\ref{ckv}) and a solution
to
the equation can be viewed as a natural extension of symmetries from
infinity. 

Four linear independent classes of vector solutions to the generalized 
almost-Killing equation (\ref{alki}) in the Kerr space-time have been 
found~\cite{Taub:1978} in terms of Teukolsky's radial and angular
functions.
The vector solutions which are asymptotic to the ten Killing vectors of
Minkowski
space-time have been also given.

Though the almost-Killing symmetry approach does not have any factor
measuring 
the degree of inhomogeneity, it is the only known approximate symmetry
approach
where the corresponding equations have been solved explicitely and the 
corresponding group has been found. The technique is global and possesses 
a clear physical interpretation. There is an explicit relation between
almost-Killing
and almost symmetry (see Section \ref{sec:as}): the almost symmetry
eigenvalue equation (\ref{eige}) reduces
to the almost-Killing and conformal almost-Killing ones for ${_m}\lambda =
0$ and 
${_m}\lambda = 1/2$, respectively. On the other hand, no generalization of 
the notion of the almost-Killing
symmetry itself, as well as solutions to other metrics, are known.  

\section{Killing-like Symmetry}
\label{sec:kls}
The main idea of the Killing-like symmetry~\cite{Tava-Zala:1994},
\cite{Zala:1999} 
is to consider the 
most general form of deviation from the Killing equations. Let us consider
the 
equation for a Killing-like vector $\xi^\alpha (x^\mu)$
\begin{equation}
\label{kls}
\xi_{\alpha ; \beta} + \xi_{\beta ; \alpha} = 2 \epsilon_{\alpha \beta}
\end{equation}
where a symmetric tensor $\epsilon_{\alpha \beta}(x^\mu)$ measures
deviation  
from the Killing symmetry. The tensor can be small in order to 
enable a continuous limit to the case $\epsilon_{\alpha \beta} \rightarrow
0$. 

The equation (\ref{kls}) covers the cases of semi-Killing, almost-Killing
and 
almost symmetries with additional equations for the tensor 
$\epsilon_{\alpha \beta}(x^\mu)$. Also covered are standard
generalizations 
of Killing symmetry
such as conformal and homothetic Killing vectors~\cite{KSMH:1980}. The
algebraic
classification of the symmetric tensor $\epsilon_{\alpha \beta}$ gives an 
invariant way to introduce a set of scalar indexes measuring the degree 
of inhomogeneity of the space-time with (\ref{kls}) compared with that
with isometries,
or even weaker symmetry, for example, conformal Killing's. For the most
general case $A_1 [111,1]$ in Segre's notation~\cite{KSMH:1980}  
$\epsilon_{\alpha \beta}$ has the form
\begin{equation}
\label{a1}
\epsilon_{\mu \nu} = \lambda g_{\mu \nu} + \rho x_{\mu} x_{\nu}
+ \sigma y_{\mu} y_{\nu} + \tau z_{\mu} z_{\nu} 
\end{equation}
where $g_{\mu\nu}$ if the space-time metric, $\lambda (x^\mu)$, $\rho
(x^\mu)$, 
$\sigma (x^\mu)$ and $\tau (x^\mu)$ 
are eigenvalues of $\epsilon_{\alpha \beta}$ and 
$\bigl\{ t^{\mu}, x^{\mu}, y^{\mu}, z^{\mu} \bigr\}$ is the 
eigentetrad\footnote{An equivalent form of (\ref{a1}) without explicitly
using a metric tensor is 
$\epsilon_{\mu \nu} = - \lambda t_{\mu} t_{\nu} + ( \rho + \lambda)
x_{\mu} x_{\nu}
+ (\sigma + \lambda) y_{\mu} y_{\nu} + (\tau + \lambda) z_{\mu} z_{\nu}$.
}. 
If all eigenvalues vanish the space-time has an isometry (\ref{kls}),
if $\rho = \sigma = \tau =0$ then there is a conformal Killing vector 
for $\lambda (x) \neq 0$
and a homothetic Killing vector for $\lambda = {\rm const}$. For other
algebraic types
of Killing-like symmetry the space-time has the following sets of
eigenvalues:
two complex conjugated to each other and two real scalars for $A_2
[11,ZZ^*]$, 
three real scalars for $A_3 [11,2]$ and two real scalars for $B [1,3]$.

This approach enables one to control the situation under averaging to see
how 
an approximate symmetry gives rise to `more' precise one, or remains the
same 
if a space-time initially was enough symmetric (see
Section~\ref{sec:ias}).
Under a covariant space-time averaging 
scheme~\cite{Zala:1992}, \cite{Zala:1993}, \cite{Zala:1997} generalizing 
the standard Minkowski space-time averaging scheme~\cite{Mars-Zala:1997},
a Killing vector 
(\ref{kill}) has been shown to remain a Killing vector in the averaged
space-time
\begin{equation}
\label{avkv}
\overline\xi_{\alpha \parallel \beta} + \overline\xi_{\beta \parallel
\alpha} = 0
\end{equation}
where $\parallel$ is the covariant derivative with respect to macroscopic 
metric~\cite{Zala:1992}, \cite{Zala:1993}, \cite{Zala:1997} $G_{\alpha
\beta}$. 
In contrast to a smooth and highly symmetric macroscopic space-time, a
microscopic
space-time having plenty of small-scale inhomogeneities cannot be expected
in general 
to possess isometries. Assuming that approximate microscopic symmetry
belongs 
to the class of Killing-like symmetry (\ref{kls}), one can
show~\cite{Tava-Zala:1994}, 
\cite{Zala:1999} that averaged Killing-like equation reads
\begin{equation}
\label{avkl}
\overline \xi_{\alpha \parallel \beta} + \overline \xi_{\beta \parallel
\alpha} 
= 2 \overline \epsilon_{\alpha \beta}.
\end{equation}
Given a microscopic Killing-like symmetry, for example, $A_1 [111,1]$ with 
(\ref{a1}), the average tensor $\overline \epsilon$ can be shown to have
the following
form: 
\begin{equation}
\label{ava1}
\overline \epsilon_{\mu \nu} = \langle \lambda \rangle \overline g_{\mu
\nu}
+ \langle \rho \rangle X_{\mu} X_{\nu} + \langle \sigma \rangle Y_{\mu}
Y_{\nu} 
+ \langle \tau \rangle Z_{\mu} Z_{\nu}
\end{equation}
where $\langle \lambda \rangle (x^\mu)$, $\langle \rho \rangle (x^\mu)$, 
$\langle \sigma \rangle (x^\mu)$ and $\langle \tau \rangle (x^\mu)$ are 
rational functions of eigenvalues of 
averaged metric $\overline g_{\mu \nu}$, averaged symmetric tensors
$\overline{t_{\mu} t_{\nu}}$, $\overline{x_{\mu} x_{\nu}}$,
$\overline{y_{\mu} y_{\nu}}$ 
and $\overline{z_{\mu} z_{\nu}}$ and of averaged eigenvalues of 
(\ref{a1}), $\overline \lambda$, $\overline \rho$, $\overline \sigma$ and
$\overline \tau$,
with the {\em linear} dependence on the last-mentioned,  
and $\bigl\{ T^{\mu}, X^{\mu}, Y^{\mu}, Z^{\mu} \bigr\}$ is the 
macroscopic eigentetrad in the averaged (macroscopic) space-time.
The correlations between eigenvalues and eigenvectors 
of (\ref{a1}) are taken into account by algebraic decomposition of 
symmetric second rank correlation tensors with the correlation eigenvalues 
entering the quantities $\langle \lambda \rangle (x^\mu)$, $\langle \rho
\rangle (x^\mu)$, 
$\langle \sigma \rangle (x^\mu)$ and $\langle \tau \rangle (x^\mu)$ in
(\ref{ava1}) as
additive terms. Without loss of generality one can consider the
correlation terms 
renormalizing the averaged eigenvalues $\overline \lambda$, $\overline
\rho$, 
$\overline \sigma$ and $\overline \tau$. The only assumption made in
derivation of 
(\ref{ava1}) is that averaging out {\em does not change} the algebraic
type,
that is the averaging of the types $A_1 [1,111]$
$A_2 [11,ZZ^*]$, $A_3 [11,2]$ and $B [1,3]$ may lead at most to
degeneration
of the same type, not allowing interchanges between the types
\footnote{This is dictated by the local character of both the algebraic 
decomposition and the averaging procedure.}.

If upon averaging $\overline \lambda =0$, $\overline \rho = 0$, 
$\overline \sigma = 0$ and $\overline \tau = 0$, i.e. the original
space-time
had no symmetries (was absolutely disordered), then the microscopic 
$\xi_{(\alpha ; \beta)} = \epsilon_{\alpha \beta}$ 
leads to a Killing symmetry $\overline \xi_{( \alpha \parallel \beta)} =
0$
in the macroscopic space-time. If $\overline \lambda \neq 0$, 
$\overline \rho = \overline \sigma = \overline \tau = 0$ then there is 
a Killing-like symmetry $\overline \xi_{( \alpha \parallel \beta)} 
=  \overline \lambda \overline g_{\alpha \beta}$ which reduces to 
a conformal Killing symmetry $\overline \xi_{( \alpha \parallel \beta)} 
=  \overline \lambda (x^\mu) G_{\alpha \beta}$ only when there are no 
correlations in eigenvectors and all eigenvalues of $\overline g_{\alpha
\beta}$ 
are equal to unity, $\overline{t_{\mu} t_{\nu}} = T_{\mu} T_{\nu}$, 
$\overline g_{\alpha \beta} T^{\beta} = T_{\alpha}$ etc.

Let us consider now a particular case of the microscopic Killing-like
symmetry taken 
as a homothetic Killing vector with the homothetic parameter $\lambda =
1$. Then
the microscopic equation $\xi_{(\alpha ; \beta)} =  g_{\alpha \beta}$
leads to
the macroscopic Killing-like symmetry:
\begin{equation}
\label{homo}
\overline \xi_{(\alpha \parallel \beta)} =  \overline g_{\alpha \beta}.
\end{equation}
By means of analysis of the integrability 
conditions\footnote{In macroscopic gravity the metric tensor and its 
inverse remain covariantly constant under averaging, 
$\overline g_{\alpha \beta \parallel \gamma} = 0$, 
$\overline g^{\alpha \beta}{}_{\parallel \gamma} = 0$, 
$g_{\alpha \beta} g^{\alpha \gamma} \neq \delta_\beta^\gamma$, 
with the covariantly constant macroscopic metric 
$G_{\alpha \beta \parallel \gamma} = 0$ and
its inverse $G^{\alpha \beta}{}_{\parallel \gamma} = 0$, 
$G_{\alpha \beta} G^{\alpha \gamma} = \delta_\beta^\gamma$.} of
(\ref{homo}) 
one can prove a theorem classifying all space-times with the Killing-like 
symmetry (\ref{homo}). 
\vskip 0.2cm
{\bf Theorem} \, Depending on algebraic types of $\overline g^{\alpha
\beta}$,
the following space-times admit the Killing-like symmetry (\ref{homo}):

(A) All algebraic types $A_1 [111,1]$, $A_2 [11,ZZ^*]$, 
$A_3 [11,2]$ and $B [1,3]$  
without degeneration have 4 linear independent
covariantly constant vector fields, i.e. the macroscopic space-time is
flat.

(B) For $A_3 [(11,2)]$ when 
\begin{equation}
\label{a3}
\overline g_{\alpha \beta} = {\rm const} \, G_{\alpha \beta} - 2
\sigma_{\alpha} \sigma_{\beta}
\end{equation}
(a generalized homothetic Killing vector)
the space-time has a covariantly constant null vector
$\sigma^{\alpha}{}_{\parallel \beta} = 0$,
i.e. $G_{\alpha \beta}$ is a $pp-$wave metric~\cite{KSMH:1980}.

(C) For $A_3 [(111,1)]$ when 
\begin{equation}
\label{a1de}
\overline g_{\alpha \beta} = {\rm const} \, G_{\alpha \beta} 
\end{equation}
(a homothetic Killing vector)
there are no conditions on the curvature tensor and space-time structure.

(D) All other cases are given in Table \ref{theo}.
\vskip 0.2cm

An important advantage of the Killing-like symmetry approach is that it
incorporates
all known geometrical approaches, semi-Killing, almost-Killing and 
almost symmetries. A link to the approximate symmetry group approach is
not clear
as yet, and a group picture of the Killing-like symmetry is not available. 
The technique is essentially local due to the local character of algebraic
decomposition, though under certain circumstances, such as the existence
of 
non-singular vector fields, it may be extended on the entire
manifold. Another advantage is its applicability to metrics of any 
signature. On the other hand, the physical interpretation of the
Killing-like 
symmetries and of the eigenvalues (\ref{kls}) as candidates for
inhomogeneity 
indexes is not known so far.   

\section*{Acknowledgments}
The author would like to thank Remo Ruffini for hospitality in ICRA where 
the work has been done in part. Discussions with Reza Tavakol, George 
Ellis, Vahe Gurzadyan and Giovanni Montani are gratefully acknowledged.

\newpage

\begin{table}[t]
\begin{center}
\footnotesize
\caption{Part (D) of the Theorem in Section \ref{sec:kls} includes the 
following cases (here all vectors and symmetric idempotent second rank
tensors are covariantly constant, vanishing eigenvalues are marked by zero
and
asterisk stands for complex conjugate):
\label{theo}}
\begin{tabular}{|l|l|c|c|} \hline
tensor $\overline g_{\alpha \beta}$ & Ricci tensor & Petrov type &
Space-time admits \\ \hline
$[11(1,1)]$ & $[(\stackrel{0}{1}\stackrel{0}{1})(1,1)]$ & $D$ & two
spacelike vectors 
                                                                 \& one
tensor \\ \hline
$[(11)1,1]$ & $[(11)(\stackrel{0}{1},\stackrel{0}{1})]$ & $D$ & spacelike
\& timelike vectors 
\& one tensor \\ \hline
$[(11)(1,1)]$ & $[(11)(1,1)]$       & $D$ & two tensors \\ \hline
$[1(11,1)]$ &                       &     & one spacelike vector \& one
tensor \\ 
            & $[\stackrel{0}{1}11,1]$ & $I$ &
\\
            & $[\stackrel{0}{1}1(1,1)]$ & $D$ &
\\
            & $[\stackrel{0}{1}(11),1]$ & $D$ &
\\
            & $[\stackrel{0}{1}(11,1)]$ & $O$ &
\\
            & $[\stackrel{0}{1}1,ZZ^*)]$ & $I$ &
\\
            & $[\stackrel{0}{1}1,2]$     & $II$ &
\\
            & $[\stackrel{0}{1}(1,2)]$   & $N$ &
\\
            & $[\stackrel{0}{1},3]$      & $III$ &
\\ \hline
$[(111),1]$ &                        &    &  one timelike vector \& one
tensor \\
            & $[111,\stackrel{0}{1}]$    & $I$ &
\\ 
            & $[(11)1,\stackrel{0}{1}]$    & $D$ &
\\
            & $[(111),\stackrel{0}{1}]$    & $O$ &
\\ \hline
$[(11),ZZ^*]$ & $[(11)(\stackrel{0}{1},\stackrel{0}{1})]$ & $D$ &
spacelike \& timelike
                                                              vectors \&
one tensor \\ \hline
$[1(1,2)]$  & $[(\stackrel{0}{1} \stackrel{0}{1}, \stackrel{0}{2})]$ & $N$
& spacelike \&

null vectors \\ \hline
$[(11),2]$  & $[(11)(\stackrel{0}{1}, \stackrel{0}{1})]$ & $D$ & spacelike
\& timelike vectors
                                                              \& one
tensor \\ \hline
$[(1,3)]$   & $[(\stackrel{0}{1} \stackrel{0}{1}, \stackrel{0}{2})]$ & $N$
& spacelike \&

null vectors \\ \hline
\end{tabular}
\end{center}
\end{table}


\begin{thebibliography}{99}
\bibitem{Beke:1973} J. Bekenstein, {\em Phys. Rev. D} {\bf 7} (1973)
{2333}.

\bibitem{Hawk:1975} S. Hawking, {\em Commun. Math. Phys.} {\bf43} (1975)
{199}.

\bibitem{Penr:1979}R. Penrose in {\em General Relativity}, eds. S.W.
Hawking and
W. Israel (Cambridge University Press, Cambridge, 1979), p. 

\bibitem{Hu:1983} B.L. Hu, {\em Phys. Lett.} {\bf 97A} (1983) {368}.

\bibitem{BMP:1993} R. Blanderberger, V. Mukhanov and T. Prokopec, 
{\em Phys. Rev. D} {\bf 48} (1993) {2443}.

\bibitem{Smol:1985} L. Smolin, {\em Gen. Relat. Grav.} {\bf 17} (1985)
{417}.

\bibitem{Beke:1994} J. Bekenstein, Do we understand black hole entropy?
{\em gr-qc/9409015} (1994).

\bibitem{Davi:1978} P.C.W. Davies, {\em Rep. Prog. Phys.} {\bf 41} (1978)
{1313}.

\bibitem{Hake:1978} H. Haken, {\em Synergetics} (Springer-Verlag, Berlin,
1978).

\bibitem{Shan:1948} C.E. Shannon, {\em Bell System Tech. J.} {\bf 27}
(1948) {379},
{\em ibid.} {\bf 27} (1948) {623}; 
reprinted in C.E. Shannon and W. Weaver, {\em The Mathematical Theory of
Communications} 
(University of Illinois Press, Urbana, 1949).

\bibitem{Bril:1956} L. Brillouin, {\em Science and Information Theory}
(Academic Press,
New York, 1956).

\bibitem{Jayn:1957a} E.T. Jaynes, {\em Phys. Rev.} {\bf 106} (1957) {620}.

\bibitem{Jayn:1957b} E.T. Jaynes, {\em Phys. Rev.} {\bf 108} (1957) {171}.

\bibitem{Matz:1968a} R.A. Matzner, {\em J. Math. Phys.} {\bf 9} (1968)
{1063}.

\bibitem{Mont-Zala:1999} G. Montani and R.M. Zalaletdinov  
in {\em Proc. of the 8th Marcel Grossmann Meeting on General Relativity}, 
Part A, ed. T. Piran (World Scientific, Singapore, 1999), p. 628.

\bibitem{Matz:1968b} R.A. Matzner, {\em J. Math. Phys.} {\bf 9} (1968)
{1657}.

\bibitem{Koma:1962} A. Komar, {\em Phys. Rev.} {\bf 127} (1962) {1411}.

\bibitem{Koma:1963} A. Komar, {\em Phys. Rev.} {\bf 129} (1963) {1873}.

\bibitem{Elli:1984} G.F.R. Ellis, in {\em General Relativity and
Gravitation},
eds. B. Bertotti, F. de Felice and A. Pascolini (D. Reidel Publishing
Company, 
Dordrecht, 1984), p. 215.

\bibitem{Tava-Zala:1994} R. Tavakol and R. Zalaletdinov, unpublished
(1994).

\bibitem{Zala:1999} R. Zalaletdinov, Approximate Symmetries, Inhomogeneous
Spaces and Gravitational Entropy, {\em Proc. of the Second ICRA Network
Workshop ``The Chaotic Universe"}, eds. V. Gurzadyan and R. Ruffini (World
Scientific, 
Singapore, 2000), to appear.

\bibitem{Sper-Baie:1977} A. Spero and R. Baierlein, {\em J. Math.
Phys.} {\bf 18} (1977) {1330}.

\bibitem{Sper-Baie:1978} A. Spero and R. Baierlein, {\em J. Math. 
Phys.} {\bf 19} (1978) {1324}.

\bibitem{Trau:1962} A. Trautman, in {\em Gravitation: An Introduction 
to Current Research}, ed. L. Witten (John Wiley \& Sons, New York, 1962);
also 
{\em Lectures on Relativity}, King's College, London, unpublished (1958).

\bibitem{Pere:comm} A. Peres, private communication from C.W. Misner;
cited 
according to A. Komar, {\em Phys. Rev.} {\bf 129} (1963) {1873}.

\bibitem{Yano-Boch:1953} K. Yano and S. Bochner, {\em Curvature and Betti
Numbers}, (Princeton University Press, Princeton, 1953).

\bibitem{Isaa:1968a} R.A. Isaacson, {\em Phys. Rev.} {\bf 166} (1968)
{1263}.

\bibitem{Isaa:1968b} R.A. Isaacson, {\em Phys. Rev.} {\bf 166} (1968)
{1272}.

\bibitem{ZTE:1996} R. Zalaletdinov, R. Tavakol and G.F.R. Ellis, 
{\em Gen. Relat. Grav.} {\bf 28} (1996) {1251}.

\bibitem{MTW:1973} C.W. Misner, K.S. Thorne and J.A. Wheeler, {\em 
Gravitation} (Freeman, San Francisco, 1973).

\bibitem{Bril-Hart:1964}  D.R. Brill and J.B. Hartle, 
{\em Phys. Rev.} {\bf 135} (1964) {B271}. 

\bibitem{Zala:1996} R.M. Zalaletdinov, {\em Gen. Relat. Grav.} {\bf 28}
(1996) {953}.

\bibitem{York:1974} J.W. York, {\em Ann. Inst. Henri Poincare, XXI} {\bf
4} (1974) {319}.

\bibitem{Taub:1978} C.H. Taubes, {\em J. Math. Phys.} {\bf 19} (1978)
{1515}.

\bibitem{Chan:1983} S. Chandrasekhar, {\em The Mathematical Theory of
Black Holes}
(Oxford University Press, New York, 1983).

\bibitem{Boye-Lind:1967} R.H. Boyer and R.W. Lindquist, 
{\em J. Math. Phys.} {\bf 8} (1967) {265}.

\bibitem{KSMH:1980} D. Kramer, H. Stephani, M. MacCallum and E. Herlt, 
{\em Exact Solutions of Einstein's Field Equations} (Deutsche Verlag der
Wissenschaften and Cambridge University Press, Berlin and Cambridge,
1980). 

\bibitem{Zala:1992}  R.M. Zalaletdinov, {\em Gen. Rel. Grav.} {\bf 24}
(1992) {1015}.

\bibitem{Zala:1993}  R.M. Zalaletdinov, {\em Gen. Rel. Grav.} {\bf 25}
(1993) {673}.

\bibitem{Zala:1997}  R.M. Zalaletdinov, {\em Bull. Astron. Soc. India}
{\bf 25} (1997) {401}.

\bibitem{Mars-Zala:1997}  M. Mars and R.M. Zalaletdinov, 
{\em J. Math. Phys.} {\bf 38} (1997) {4741}.

\end{thebibliography}
\end{document}